\documentclass[12pt, a4paper]{article}

\usepackage[hmargin=2.54cm,vmargin=2.54cm]{geometry}
 \usepackage{graphicx}
 \usepackage{amsmath}
 \usepackage{amssymb}
  \usepackage{enumerate}
\usepackage{cite}
\usepackage{mathabx}
\usepackage{url}
\newcommand{\bey}{\begin{eqnarray}}
\newcommand{\eey}{\end{eqnarray}}
\usepackage{etoolbox}
\apptocmd{\thebibliography}{\setlength{\itemsep}{1pt}}{}{}

\begin{document}

\noindent {\bf \large   Gravitational Decoherence in Deep Space Experiments}

\bigskip

\noindent Charis Anastopoulos\footnote{anastop@upatras.gr}

\noindent  Department of Physics, University of Patras, 26500 Greece

 \medskip

\noindent Miles Blencowe\footnote{Miles.P.Blencowe@dartmouth.edu}\\
\noindent  Department of Physics and Astronomy, Dartmouth College, Hanover, New Hampshire 03755, USA.

\medskip

\noindent Bei-Lok Hu\footnote{blhu@umd.edu}\\
\noindent Maryland Center for Fundamental Physics and Joint Quantum Institute,
 University of Maryland, College Park, Maryland 20742-4111, USA.

\bigskip

\bigskip

\noindent {\bf Abstract} 
Among the many worthwhile quantum experiments taking  advantage of long baselines in space, this white paper points to the far-reaching significance of  gravitational decoherence experiments.  These experiments can provide clues as to whether gravity is of a fundamental or an  effective nature.  They can also discriminate between  the predictions of quantum field theory in curved spacetime, our default theory for quantum phenomena in background gravitational fields, and other popular alternative quantum theories.
\\ \\
White paper   for the National Academies of Science 's Decadal Survey on Biological and Physical Sciences Research in Space 2023-2032.

\bigskip

\pagebreak

\section{Executive summary}

1.  Quantum gravity (Q$\times$G) research, namely, theories for the microscopic constituents of spacetime, operative at the Planck length-scale $L_p = 1.6 \times 10^{-35}~{\mathrm{m}}$, has occupied the attention of a significant number of theoretical physicists for the past seven decades; the lack of direct experimental data has prevented any of the resulting theories to claim success.  However, the contradictions and inconsistencies between  quantum theory and general relativity (GR), the two well tested theories for the microscopic and the macroscopic worlds,  already show up acutely at today's accessible low energy scales.  The time is ripe to probe in a different direction, namely, Q+G.

2.   Quantum Field Theory (QFT) in Curved Spacetime (CST) is the most reliable theory for the union (Q+G), not the fusion (Q$\times$G), of  QFT for matter,  and GR for gravity.  Established in the 1970's, it predicts effects like Hawking radiation from black holes \cite{hawking}, and---with extension to semi-classical gravity---it provides the theoretical framework for  inflationary cosmology \cite{Guth81} which foretells a spatially-flat universe.  Theoretical work since then has focused on strong-field quantum processes near  black holes  and in the early universe, preparing us well for cosmological  experiments and observations in the coming decades.

3. Tests of QFTCST  in earth-bound laboratories have been carried out, but so far only for {\em analog} gravity models \cite{AnalogG}; these tests are, at best, indirect.   Deep space experiments  \cite{DSQL,expdec} can provide the first {\em direct} tests of QFTCST.
Among the many worthwhile experiments taking  advantage of long baselines in space, this white paper points to the far-reaching significance of  gravitational decoherence experiments.  Environmental influences can overshadow the weak gravitational effects,   thus a space environment is ideal as it suppresses such influences. Gravitational decoherence experiments can provide clues as to whether gravity is of a fundamental or an  effective nature.  They can also discriminate between  the predictions of QFTCST and other popular alternative quantum theories.

\section{Fundamental-Theory Capabilities of Space Experiments}

Space experiments allow us to probe a different regime of interplay between QFT and GR, a regime where quantum systems extend or propagate over distances at which gravitational phenomena become important. In particular, deep space experiments may involve interferometry with arms of the order of $10^5~{\mathrm{km}}$ and also the construction of entangled photon states that extend to the same distances.

\subsection{Testing whether gravity is fundamental or emergent}

As  shown in Ref.  \cite{AHGravDec}, an important feature of gravitational decoherence is that the decoherence rate depends on extra parameters other than the Planck scale. This is similar to the dependence of the decoherence rate of a quantum Brownian particle to the temperature and spectral density of the environment it interacts with. The corresponding features when gravity acts as an environment in decohering quantum objects are what Anastopoulos and Hu call the {\em textures} of spacetime. There is a marked difference between the case when gravity is represented as a background spacetime versus the case when gravity acts like a thermodynamic bath to quantum particles. This points to the possibility of using gravitational decoherence measurements to discern whether gravity is intrinsically elemental or emergent.

\subsection{Testing  alternative quantum theories}

Asserting that the suppression of interference is a consequence of decoherence mechanisms, there are several alternative quantum theories with predictions markedly different from QFTCST; we name four types in the following.  For example,  in the continuous spontaneous localization (CSL) models described below, the desired decoherence is in the configuration space while from GR, decoherence can only  happen in the energy basis.
\begin{enumerate}
\item CSL models, which postulate an {\em ad hoc} fundamental dynamical reduction mechanism as a solution to the measurement problem and the emergence of the classical world \cite{CSL}.
    \item Models that postulate gravity to be a fundamentally classical channel of interaction, thereby causing decoherence or localization of quantum particles. They include the popular Diosi-Penrose decoherence theory \cite{DP} and the Newton-Schr\"odinger equation for quantum particle dynamics \cite{NS}.
        \item Models that postulate (highly conjectural) quantum gravity effects surviving at low energies and leading to non-unitary time evolution for matter fields---see, for example, Ref. \cite{QGeff}.
            \item Models of `fundamental' decoherence that postulate stochastic fluctuations of  spacetime at the Planck-scale level---see, for example, Ref. \cite{fluctdec}.
\end{enumerate}
For tests of the models above, most experiments presuppose preparation of sufficiently heavy particles in superposition states; the particle masses  for which the postulated  decoherence phenomena are distinguishable are of the order of $10^9-10^{10}~{\mathrm{amu}}$. For a comparison,  the heaviest molecules used to date in quantum mechanical interference experiments are oligoporphyrines with a mass of ``only'' $2.6 \cdot 10^4\,~{\mathrm{amu}}$ \cite{Fein19}. However, the use of masses superpositions up to $10^{13}~{\mathrm{amu}}$ have been predicted for superconducting, magnetically-levitated microspheres \cite{PRI}.

All of the models listed above predict a violation of QFT, indeed, perhaps just for the sake of being different, most models explicitly seek such a violation. It turns out that most models are also incompatible with GR, in the sense that they violate the fundamental symmetry of GR, namely, invariance under spacetime diffeomorphisms \cite{AHGravDec} (often referred to as the principle of General Covariance). Since GR and QFT are fundamental theories that have passed strong tests in many different regimes,  theories contradicting both bring about enormous stakes in experimental tests, but have  a low {\em a priori} probability to be true, especially since they make very strong theoretical assumptions \cite{AnHucr}.

\section{The ABH model}

A theory of gravitational decoherence  based on QFT and GR does exist, that which was originally proposed by  Anastopoulos \& Hu \cite{AHGravDec} and Blencowe \cite{Miles}  (ABH). (See also the later Ref. \cite{Oniga}, which derives a related theory of gravitational decoherence for matter and light.)    In this theory, the source of decoherence comes  from the noise  (fluctuations) in gravitational waves which are the transverse-traceless perturbations,  or gravitons (quantized linear perturbations) \cite{graviton}.  The source of these fluctuations may be cosmological, astrophysical,  or structural to GR as an emergent theory.

GR is commonly accepted as the best theory for the description of macroscopic spacetime, but whether quantizing GR will yield the true theory of the microscopic structure of spacetime  at the Planck scale remains an open question.  GR could well be  an emergent theory from some fundamental theory of  quantum gravity,   valid only at the macroscopic scale we are familiar with.  The key benefit of the ABH model is that it can distinguish these two alternatives.  In the fundamental  theory view,  Minkowski spacetime is  the ground state of a quantum gravity theory.  In the emergent theory view, Minkowski spacetime is a low energy collective state or  {\em macrostate} of quantum gravity, whereby one could associate a thermodynamic  description. The key difference between a ground state and a macrostate is not energy,  but the {\em strength of fluctuations}: thermodynamic fluctuations are much stronger than quantum  fluctuations in spacetime, and they can cause significantly stronger decoherence.

 The ABH model involves a single free parameter, the noise temperature $\Theta$, which characterizes the strength of gravitational fluctuations. For $\Theta = 0$, the fluctuations correspond to the quantum vacuum, while any non-zero value of $\Theta$ describes thermodynamic fluctuations.
The parameter $\Theta$    conveys coarse-grained information reflective of the micro-structures of spacetime, similar to temperature with regard to molecular motion.   It is in this sense that gravitational decoherence may reveal the underlying textures of spacetime beneath that described by classical GR.

Since $\Theta$ is a free parameter that measures the strength of noise and not an actual temperature, it is not constrained to be smaller than the Planck temperature $T_P = 1.4\times 10^{32}~{\mathrm{K}}$. In principle, $\Theta$ could be much larger than $T_P$; however, cosmological considerations suggest that $\Theta$ must be at most of the order of $T_P$. Solar system physics also suggests an upper bound to  $\Theta$ of a few orders of magnitude  smaller than $T_P$. An exact bound requires an analysis of the motion of planet-sized bodies under ABH-type noise.

For free non-relativistic particles of mass $m$, the ABH model predicts a master equation
\begin{eqnarray}
 \frac{\partial \hat{\rho}}{\partial t} = -i [\hat{H}, \hat{\rho}] - \frac{\tau_{NR}}{16m^2} (\delta^{ij} \delta^{kl} + \delta^{ik}\delta^{jl}) [\hat{p}_i
\hat{p}_j,[\hat{p}_k\hat{p}_l, \hat{\rho}]]
\end{eqnarray}
 where
  \begin{eqnarray}
 \tau_{NR} = \frac{32\pi G \Theta}{9} = \frac{32\pi}{9} \tau_P (\Theta/T_P),
 \end{eqnarray}
 is a constant with the dimensions of time and $\hat{H} = \frac{\hat{p}^2}{2m}$.

 For motion in one dimension, the ABH master equation simplifies to
 \begin{eqnarray}
 \frac{\partial \hat{\rho}}{\partial t} = -i [\hat{H}, \hat{\rho}] - \frac{\tau_{NR}}{2} [\hat{H},[\hat{H},\hat{\rho}]]. \label{1dim}
 \end{eqnarray}

The ABH model can be generalized to photons, where we can obtain a master equation for a general photon state \cite{LagAn}. For a single photon,
\begin{eqnarray}\label{result}
 \frac{\partial \hat{\rho}}{\partial t} &= -i \, [ \hat{H } , \hat{\rho}] - \frac{\tau_{ph}}{2}   \left( \delta^{in} \delta^{jm} -\frac{1}{3} \delta^{ij} \delta^{nm} \right) \left[ \frac{\hat{p}_i \, \hat{p}_j}{\hat{p}_0} , \left[ \frac{\hat{p}_n \, \hat{p}_m}{\hat{p}_0}, \hat{\rho}  \right] \right] \, ,
\end{eqnarray}
where $\hat{H} = |\hat{\bf p}|$ and $\tau_{ph} = 4 G \Theta$.

\section{Observational implications of the ABH model}
\begin{enumerate}
\item {\em Optomechanical setups.} Consider a body brought into a superposition of a zero momentum and a finite momentum state, corresponding to an energy difference $\Delta E$. For the ABH model, the decoherence rate for the center of mass is then
     \begin{eqnarray}
     \Gamma_{ABH} = \frac{ (\Delta E)^2 \tau_{NR}}{\hbar^2}.
     \end{eqnarray}
    A value for $\Gamma_{ABH}$ of the order of $10^{-3}~{\mathrm{s}}$ may be observable in optomechanical systems, since it is within a few orders of magnitude of current measured environment-induced-decoherence timescales. Hence,  to exclude values of $\tau_{NR} > \tau_P$, we must prepare a quantum state with energy superposition difference  $\Delta E  \sim 10^{-14}~{\mathrm{J}}$.

 \item {\em Interference experiments} The ABH model   leads to loss of phase coherence of the order of
 $(\Delta \Phi)^2 = m^2 v^3 \tau L/\hbar^2$, where $L$ is the propagation distance inside the interferometer.
 Setting an upper limit of $L = 100~{\mathrm{km}}$, and $v = 10^4~{\mathrm{m}}/{\mathrm{s}}$,  experiments with particles having mass $10^{10}~{\mathrm{amu}}$
 will test up to $\Theta \sim 10^{-5}T_P$.

   Crucially, the ABH model leads to decoherence in the momentum or energy basis, and not in the position basis, like other models of gravitational decoherence. The dependence of decoherence on particle mass and velocity is very different from most models. If a decoherence effect is detected, it will be a straightforward task to distinguish whether its source is ABH physics or something different.

 \item {\em Decoherence of photons.} The   ABH model is, so far, the only QFT approach that has been generalized to the gravitational decoherence of photon superposition states\cite{Oniga,LagAn}, and hence is the  only one capable of making quantitative, first principles predictions for such effects in space experiments. For interferometer experiments with arm length $L$, the ABH model predicts a loss of visibility of  the order
     \begin{eqnarray}
     (\Delta \Phi)^2 = \frac{8G\Theta E^2L}{\hbar^2 c^6}.
      \end{eqnarray}
      For $L=10^5$km, $\Theta \sim T_P$ and photon energies $E$ of the order of 1eV, this implies a loss of coherence of the order of $\Delta \Phi = 10^{-8}$.  In principle, this would be discernible with EM-field coherent states with mean photon number $\bar{N} > 10^{16}$, i.e., it would be possible with LIGO sensitivities.

The linear  dependence of $\Delta \Phi$ on energy implies that decoherence is significantly stronger at high
frequencies. For interferometry in the extreme UV, $\Delta \Phi$  may increase by two orders of
magnitude or more.
     Alternative set-ups, such as the formation of effective Fabry-Perot `cavities' with mirrors could increase the effective propagation length by many orders of magnitude, and hence, lead to stronger signatures of ABH-predicted, photon gravitational decoherence.

 \end{enumerate}
\pagebreak

\end{document}